\documentstyle[amstex,manuscript,osa]{revtex}
\begin{document}
\title{Thermal Fluctuations of Elastic Ring}
\author{Y. Rabin\cite{yit} and S. Panyukov\cite{serg}}
\address{Department of Physics, Bar--Ilan University, Ramat--Gan 52900, Israel }
\maketitle
\pacs{87.15.-v, 87.15.Ya, 05.40.-a}

\begin{abstract}
We study the effects of thermal fluctuations on a small elastic ring. We
derive analytical expressions for the correlation functions of the Euler
angles, for the real space two--point correlation functions and for the
probability distribution of writhe, as a function of the persistence lengths
that determine the rigidity with respect to bending and twist. Fluctuation
amplitudes diverge in the limit of vanishing twist rigidity. There is
crossover from a small scale regime in which twist and writhe modes are
uncoupled, to a large scale regime in which twist affects the spatial
configurations of the ring.
\end{abstract}

Small circular loops of extrachromosomal DNA (plasmids) play an important
role in biological processes such as gene transfer between bacteria and in
biothechnological applications where they are used as vectors for DNA
cloning \cite{ENCYC}. The simplest  model that captures both the topology
and the physical properties of such an object is that of an elastic ring,
and several studies of the writhing instability of twisted rings based on
this model, were reported in recent years\cite{Tobias,White}. However, these
studies focused on the elastic response of the ring to mechanical forces and
did not consider the effects of thermal fluctuations. These fluctuations
dominate the physics of macromolecules and determine all their statistical
properties, such as radii of gyration, dynamics in solution\cite{PG},
kinetics of loop formation, and dissociation of short DNA segments\cite
{Volog} and molecular beacons\cite{Albert}. Recently, we developed a theory
of fluctuating elastic filaments, with arbitrary spontaneous curvature,
torsion and twist in their stress--free state\cite{PRL2000}. Since
topological constraints were not taken into account, our analysis was
limited to linear filaments and could not be directly applied to the study
of closed objects that have the topology of a ring. In the present letter we
present the statistical mechanics of weakly fluctuating rings. We calculate
the correlation functions of the Euler angles, and use them to derive the
real space correlation functions, and the complete probability distribution
function of writhe fluctuations. This information allows us to understand
the effects of twist and bending rigidities on the statistical properties of
fluctuating elastic rings.

Consider an undeformed circular ring with a centerline that forms a circle
of radius $r$ in the $xy$ plane. The Euler angles that describe this
equilibrium configuration are $\theta _{0}=\pi /2,$ $\varphi _{0}=s$ and $%
\psi _{0}=0.$ Here and in the following, the dimensionless contour length $s$
is measured in units of $r$ and thus $0\leq s\leq 2\pi .$ In the presence of
thermal fluctuations, the Euler angles $\left\{ \eta \right\} =\left\{
\theta ,\varphi ,\psi \right\} $ deviate from their equilibrium values, $%
\left\{ \eta _{0}\right\} =\left\{ \theta _{0},\varphi _{0},\psi
_{0}\right\} $, and the instantaneous deviations are denoted by $\delta \eta
(s)=\eta (s)-\eta _{0}(s)$. The topology of the ring imposes periodic
boundary conditions on the fluctuations of the Euler angles $\delta \theta
(2\pi )=\delta \theta (0),$ $\delta \psi \left( 2\pi \right) =\delta \psi (0)
$ and $\delta \varphi (2\pi )=\delta \varphi (0)$, as well as on the
fluctuations of the ring in three dimensional space, $\delta {\bf x}(2\pi
)=\delta {\bf x}(0)$. Introducing the Fourier transforms, $\delta \eta
(s)=\sum_{n}\tilde{\eta}(n)e^{ins}$, where the sum goes over all positive
and negative integers and $\tilde{\eta}(-n)=\tilde{\eta}^{\ast }(n)$)$,$ the
above boundary conditions are expressed as $\tilde{\theta}(0)=\tilde{\varphi}%
(1)=0.$ From the general expression for the elastic energy of a filament
with arbitrary spontaneous curvature, torsion and twist\cite{White,PRL2000},
one can derive the elastic energy $U$ for small deviations of the Euler
angles from their equilibrium values, 
\begin{align}
%TCIMACRO{\dfrac{U}{2\pi T}}%
%BeginExpansion
{\displaystyle{U \over 2\pi T}}%
%EndExpansion
& =%
%TCIMACRO{\dfrac{a_{b}}{2}}%
%BeginExpansion
{\displaystyle{a_{b} \over 2}}%
%EndExpansion
\left| \tilde{\psi}(0)\right| ^{2}+\left( a_{b}+a_{t}\right) \left| i\tilde{%
\theta}(1)+\tilde{\psi}(1)\right| ^{2}  \label{energy} \\
+& \sum_{n=2}^{\infty }\left[ a_{b}\left| in\tilde{\theta}(n)+\tilde{\psi}%
(n)\right| ^{2}+a_{b}n^{2}\left| \tilde{\varphi}(n)\right| ^{2}+a_{t}\left|
in\tilde{\psi}(n)-\tilde{\theta}(n)\right| ^{2}\right] ,  \nonumber
\end{align}
where $T$ is the temperature and $a_{b}$ and $a_{t}$ are dimensionless
persistence lengths (given in units of the radius of the ring $r$)
associated with bending and twist, respectively, and are proportional to the
elastic moduli that determine the rigidity with respect to the corresponding
deformations. The quadratic approximation for the energy, Eq. (\ref{energy}%
), holds in the limit of weak fluctuations of the Euler angles, i.e., when
their characteristic fluctuation amplitude satisfies $\delta \eta _{char}\ll
1$ (in radians)$.$ This energy do not depend on modes $\tilde{\psi}(1)=-i%
\tilde{\theta}(1)$ and $\tilde{\varphi}(0)$ that correspond to rigid--body
rotation of the entire ring, with respect to axes lying in the plane of the
ring and normal to it, respectively. The expression in the square brackets
in Eq. (\ref{energy}) contains terms ($\tilde{\psi}(n),$ $\tilde{\theta}(n)$%
) that are not multiplied by $n$ and consequently depend on the values of
the corresponding angles in the stress--free state (the energy depends
explicitly on the spontaneous curvature of the filament).

The energy, Eq. (\ref{energy}), is the sum of contributions each of which is
a quadratic form that can be represented (for $n\geq 2$) as a $3\times 3$
matrix in the space spanned by $\tilde{\eta}(n)$. Since this matrix is
diagonal in $\tilde{\varphi}(n),$ with  eigenvalues $\lambda
_{1}(n)=a_{b}n^{2},$ the fluctuations of the angle $\varphi $ decouple from
those of the angles $\theta $ and $\psi .$ The energy of this mode depends
on the bending modulus only and therefore $\delta \varphi (s)$ describes
pure bending fluctuations in the $xy$ plane. The remaining $2\times 2$
matrix in $\left\{ \tilde{\theta}(n),\tilde{\psi}(n)\right\} $ space$\ $can
be diagonalized by solving for the roots of a quadratic characteristic
equation,  $\lambda _{2,3}(n)=\left( a_{b}+a_{t}\right) \left(
n^{2}+1\right) /2\mp \left[ \left( a_{b}-a_{t}\right) \left( n^{2}+1\right)
^{2}/4+4n^{2}a_{b}a_{t}\right] ^{1/2}$.

We proceed to calculate the correlation functions of the Euler angles, 
\begin{equation}
\left\langle \delta \eta \left( s\right) \delta \eta ^{\prime }\left(
0\right) \right\rangle =\sum_{n}e^{ins}\left\langle \tilde{\eta}(n)\tilde{%
\eta}^{\prime }(-n)\right\rangle =\sum_{n,k}e^{ins}%
%TCIMACRO{\dfrac{\eta _{k}(n)\eta _{k}^{\prime }(-n)}{\pi \lambda _{k}(n)}}%
%BeginExpansion
{\displaystyle{\eta _{k}(n)\eta _{k}^{\prime }(-n) \over \pi \lambda _{k}(n)}}%
%EndExpansion
,  \label{cor}
\end{equation}
where $\tilde{\eta}(n)$ and $\tilde{\eta}^{\prime }(n)$ are components of
the vector $\left\{ \tilde{\theta}(n),\tilde{\varphi}(n),\tilde{\psi}%
(n)\right\} .$ The second equality in the above equation was derived by
expanding the modes $\tilde{\eta}(n)$ in the basis of orthonormal
eigenvectors $\eta _{k}(n)$ of the quadratic form, Eq. (\ref{energy}), $%
\tilde{\eta}(n)=\sum_{k}c_{k}(n)\eta _{k}(n)$, and using the equipartition
theorem to calculate the correlator of the coefficients in this expansion, $%
\left\langle c_{k}(n)c_{k^{\prime }}(-n^{\prime })\right\rangle =\left[ \pi
\lambda _{k}(n)\right] ^{-1}\delta _{nn^{\prime }}\delta _{kk^{\prime }}$.
The modes with $n=0,\pm 1$ should be considered separately, because some of
the corresponding eigenvalues vanish. Since the mode $\delta \varphi $
decouples from $\delta \theta $ and $\delta \psi ,$ the cross--correlators
of $\delta \varphi $ with $\delta \theta $ and $\delta \psi $ vanish
identically, $\left\langle \delta \theta (s)\delta \varphi \left( 0\right)
\right\rangle =\left\langle \delta \psi (s)\delta \varphi \left( 0\right)
\right\rangle =0$. A straightforward calculation yields the following
expressions for all the other correlators, 
\begin{align}
\pi \left\langle \delta \theta (s)\delta \theta \left( 0\right)
\right\rangle & =%
%TCIMACRO{\dfrac{\cos s}{a_{b}+a_{t}}}%
%BeginExpansion
{\displaystyle{\cos s \over a_{b}+a_{t}}}%
%EndExpansion
+%
%TCIMACRO{\dfrac{1}{a_{b}}}%
%BeginExpansion
{\displaystyle{1 \over a_{b}}}%
%EndExpansion
f_{1}\left( s\right) +%
%TCIMACRO{\dfrac{1}{a_{t}}}%
%BeginExpansion
{\displaystyle{1 \over a_{t}}}%
%EndExpansion
f_{3}\left( s\right) ,  \nonumber \\
\pi \left\langle \delta \varphi (s)\delta \varphi \left( 0\right)
\right\rangle & =%
%TCIMACRO{\dfrac{1}{a_{b}}}%
%BeginExpansion
{\displaystyle{1 \over a_{b}}}%
%EndExpansion
f_{2}\left( s\right) ,  \nonumber \\
\pi \left\langle \delta \psi (s)\delta \psi (0)\right\rangle & =%
%TCIMACRO{\dfrac{1}{a_{b}}}%
%BeginExpansion
{\displaystyle{1 \over a_{b}}}%
%EndExpansion
+%
%TCIMACRO{\dfrac{\cos s}{a_{b}+a_{t}}}%
%BeginExpansion
{\displaystyle{\cos s \over a_{b}+a_{t}}}%
%EndExpansion
+%
%TCIMACRO{\dfrac{1}{a_{t}}}%
%BeginExpansion
{\displaystyle{1 \over a_{t}}}%
%EndExpansion
f_{1}\left( s\right) +%
%TCIMACRO{\dfrac{1}{a_{b}}}%
%BeginExpansion
{\displaystyle{1 \over a_{b}}}%
%EndExpansion
f_{3}\left( s\right) ,  \nonumber \\
\pi \left\langle \delta \theta (s)\delta \psi (0)\right\rangle & =-%
%TCIMACRO{\dfrac{\sin s}{a_{b}+a_{t}}}%
%BeginExpansion
{\displaystyle{\sin s \over a_{b}+a_{t}}}%
%EndExpansion
+\left( \frac{1}{a_{t}}+\frac{1}{a_{b}}\right) f_{4}\left( s\right) ,
\label{angl}
\end{align}
where the functions $f_{1}$ through $f_{4}$ are defined by 
\begin{equation}
\begin{array}{cc}
f_{1}(s)=\left[ k(s)+1/16\right] \cos s-l(s), & f_{2}(s)=2k(s)-\cos s, \\ 
f_{3}(s)=\left[ k(s)-3/16\right] \cos s+l(s)-1/2, & f_{4}(s)=\left[ k(s)+1/16%
\right] \sin s.
\end{array}
\label{fi}
\end{equation}
with $k(s)=\left( \pi -s\right) ^{2}/8-\pi ^{2}/24,$ and $l(s)=\left[ \left(
\pi -s\right) /4\right] \sin s$. Inspection of Eqs. (\ref{angl}) shows that
bare persistence length associated with twist rigidity, $a_{t}$, plays a
fundamentally important role: the fluctuations of $\delta \theta $ and of $%
\delta \psi $ and the cross--correlations between these fluctuation modes,
diverge in the limit $a_{t}\rightarrow 0$. Therefore, simplified theories of
elastic filaments with non--vanishing spontaneous curvature that do not take
into account twist rigidity, can not describe the fluctuations of an elastic
ring. This problem does not arise in the case of straight filaments (with no
spontaneous curvature), that can be successfully described by a wormlike
chain model, in which only bending rigidity is accounted for ($a_{t}=0)$.

In Fig. 1 we plot the correlation functions of the Euler angles vs. the
dimensionless contour distance $s$, in the interval $0<s<2\pi $ (for $%
a_{b}=a_{t}=10$). Although the diagonal correlations ($\theta \theta $, $%
\varphi \varphi $ and $\psi \psi $) decay rapidly with $x$ in the range $%
s\ll 1$, the decay is not monotonic and maxima appear at $s=0,$ $\pi $ and $%
2\pi $ (these functions are symmetric with respect to reflection about $%
s=\pi $). For $\theta \theta $ and $\varphi \varphi $ correlation functions,
in between these maxima there are intervals where fluctuations of Euler
angles at points $s$ and $s^{\prime }$ are in opposition, and nodes where
these fluctuations are not correlated. This  behavior is a direct
consequence of the ring topology, and the oscillatory patterns can be
thought of as standing waves produced by interference of two wave packets
propagating along two opposite directions along the ring. The nondiagonal
correlation function $\theta \psi $ is antisymmetric with respect to
reflection about $x=\pi .$ Surprisingly,  $\theta \psi $ correlations vanish
in the limit $s\rightarrow 0,$ i.e., the fluctuations of twist and
out--of--plane deviations at two neighboring points are uncorrelated. The
underlying physics is simple: a short segment of the ring behaves as a
straight rod, for which twist about the centerline is independent of its
orientation with respect to the $xy$ plane.

The real space two--point correlation function, $\left\langle \left[ {\bf x}%
(s)-{\bf x}\left( 0\right) \right] ^{2}\right\rangle $, can be expressed in
terms of the correlation functions of the\ Euler angles. For small $s$ the
correlator depends only on the bending persistence length, and coincides
with that of a straight rod described by the wormlike chain model. In Fig. 2
we plot $\left\langle \left[ {\bf x}(s)-{\bf x}\left( 0\right) \right]
^{2}\right\rangle $ as a function of $s,$ for several choices of the bending
and twist rigidities. As expected, the mean square separation increases
parabolically with $s$ (for small $s)$ and exhibits a maximum at $s=\pi r$
(determined by the geometry of the undeformed ring). Fluctuations suppress
this maximum in a way that depends on the  rigidity parameters: decreasing
the twist rigidity $a_{t}$ has a much weaker effect than decreasing the
bending rigidity $a_{b}.$ 

We  proceed to discuss the twist and writhe fluctuations of the ring. The
twist ($Tw$) associated with a particular configuration of the ring, can be
expressed in terms of the Euler angles, 
\begin{equation}
Tw=%
%TCIMACRO{\dfrac{1}{2\pi }}%
%BeginExpansion
{\displaystyle{1 \over 2\pi }}%
%EndExpansion
\oint \omega _{3}(s)ds=%
%TCIMACRO{\dfrac{1}{2\pi }}%
%BeginExpansion
{\displaystyle{1 \over 2\pi }}%
%EndExpansion
\int_{0}^{2\pi r}\left( 
%TCIMACRO{\dfrac{d\psi }{ds}}%
%BeginExpansion
{\displaystyle{d\psi  \over ds}}%
%EndExpansion
+\cos \theta 
%TCIMACRO{\dfrac{d\varphi }{ds}}%
%BeginExpansion
{\displaystyle{d\varphi  \over ds}}%
%EndExpansion
\right) ds,  \label{Tw}
\end{equation}
where $\omega _{3}(s)$ is the ``angular velocity'' of twist about the
tangent vector\cite{PRL2000}. The writhe ($Wr$) of a given configuration of
the ring is defined by\cite{Fuller}, 
\begin{equation}
Wr=%
%TCIMACRO{\dfrac{1}{2\pi }}%
%BeginExpansion
{\displaystyle{1 \over 2\pi }}%
%EndExpansion
\oint 
%TCIMACRO{
%\dfrac{\left( {\bf e_{z}\times t}\right) \bullet \dfrac{d{\bf t}}{ds}}{1+{\bf t}\bullet {\bf e}_{z}}}%
%BeginExpansion
{\displaystyle{\left( {\bf e_{z}\times t}\right) \bullet %
{\displaystyle{d{\bf t} \over ds}} \over 1+{\bf t}\bullet {\bf e}_{z}}}%
%EndExpansion
ds,  \label{wr1}
\end{equation}
where $\times $ and $\bullet $ denote vector and scalar products,
respectively, ${\bf t}$ is a unit tangent to the ring contour, and ${\bf e}%
_{z}$ is a unit vector in $z-$direction. Expressing ${\bf t}$ in terms of
the Euler angles\cite{PRL2000} in the above expression yields 
\begin{equation}
Wr=%
%TCIMACRO{\dfrac{1}{2\pi }}%
%BeginExpansion
{\displaystyle{1 \over 2\pi }}%
%EndExpansion
\int_{0}^{2\pi r}\left( 1-\cos \theta \right) 
%TCIMACRO{\dfrac{d\varphi }{ds}}%
%BeginExpansion
{\displaystyle{d\varphi  \over ds}}%
%EndExpansion
ds.  \label{Wr}
\end{equation}

Inspection of Eqs. (\ref{Tw}), (\ref{Wr}) shows that the total rotation is
characterized by the topologically conserved  number, $Lk=Tw+Wr$ that does
not depend on the conformation of the ring\cite{Topology}. In equilibrium,
Eqs. (\ref{Tw}) and (\ref{Wr}) give $Tw_{eq}=0$ and $Wr_{eq}=1$ and although
fluctuations lead to deviations from these values, the total angle of
rotation is conserved, $2\pi Lk=2\pi Lk_{eq}=2\pi $. Expanding the Euler
angles inside the integrals in the deviations from their equilibrium values,
we obtain the following expressions for writhe and twist fluctuations, 
\begin{equation}
\delta Wr=-\delta Tw=%
%TCIMACRO{\dfrac{1}{2\pi }}%
%BeginExpansion
{\displaystyle{1 \over 2\pi }}%
%EndExpansion
\int_{0}^{2\pi r}%
%TCIMACRO{\dfrac{d\delta \varphi }{ds}}%
%BeginExpansion
{\displaystyle{d\delta \varphi  \over ds}}%
%EndExpansion
\delta \theta ds=\sum_{n}in\tilde{\varphi}(n)\tilde{\theta}(-n),  \label{dWr}
\end{equation}
where we used $\oint d\delta \varphi =\oint \delta \theta ds=0$. Since both $%
\delta \varphi $ and $\delta \theta $ enter Eq. (\ref{dWr}), $\delta Wr$
vanishes both when the fluctuations are confined to the plane of the ring ($%
\delta \theta =0$) and when they are normal to it ($\delta \varphi =0$).

Since the average $\left\langle \tilde{\varphi}(n)\tilde{\theta}%
(-n)\right\rangle $ is an even function of $n$ and the sum in Eq. (\ref{dWr}%
) goes over both positive and negative values of $n$, we find that $%
\left\langle \delta Wr\right\rangle =0$, i.e,  fluctuations do not affect
mean writhe$.$ The dispersion of the writhe is given by $\left\langle \delta
Wr^{2}\right\rangle =\sum_{n\neq 0,\pm 1}Wr^{2}(n),$ where $Wr(n)=n\left[
\left\langle \tilde{\theta}(n)\tilde{\theta}(-n)\right\rangle \left\langle 
\tilde{\varphi}(n)\tilde{\varphi}(-n)\right\rangle -\left\langle \tilde{%
\varphi}(n)\tilde{\theta}(-n)\right\rangle ^{2}\right] ^{1/2}$ is the rms
amplitude of writhe fluctuations at wavelength $r/n$. Using Eqs. (\ref{angl}%
) for the correlation functions of the Euler angles, we find 
\begin{equation}
Wr^{2}(n)=%
%TCIMACRO{\dfrac{1}{\pi ^{2}a_{b}^{2}a_{t}}}%
%BeginExpansion
{\displaystyle{1 \over \pi ^{2}a_{b}^{2}a_{t}}}%
%EndExpansion
%TCIMACRO{\dfrac{a_{b}+a_{t}n^{2}}{\left( n^{2}-1\right) ^{2}}}%
%BeginExpansion
{\displaystyle{a_{b}+a_{t}n^{2} \over \left( n^{2}-1\right) ^{2}}}%
%EndExpansion
.  \label{Wrn}
\end{equation}
For small wavelengths, $n\gg 1$, the amplitude of writhe fluctuations
depends only on the bending persistence length $a_{b}$, similarly to the
behavior of filaments that fluctuate about a straight stress--free
configuration\cite{Mezard}. Indeed, on sufficiently small scales, the ring
topology is unimportnt and the filament behaves as a straight rod whose ends
can freely rotate about the tangent direction. Since for rod with zero
spontaneous curvature, twist and writhe are not coupled through energy (via
the dependence of the elastic energy on spontaneous curvature, see Eq. (\ref
{energy})) or through topology, writhe fluctuations are independent of
twist. The crossover to the long wavelength regime at which twist and writhe
become coupled, takes place at a length scale $\xi _{t}=r\sqrt{a_{t}/a_{b}}$
and, therefore, such a regime exists in a ring of radius $r$ only if $%
a_{t}/a_{b}\leq 1.$\ The straight rod case follows from the above expression
by substituting $r=\infty ,$ and since $\xi _{t}$ diverges in this limit, we
conclude that writhe is not affected by twist, independent of the magnitude
of the twist rigidity, $a_{t}$. 

Substituting Eq. (\ref{Wrn}) back into the expression for $\left\langle
\delta Wr^{2}\right\rangle $ yields $\left\langle \delta Wr^{2}\right\rangle
=0.179a_{b}^{-2}+0.027a_{b}^{-1}a_{t}^{-1}.$ Recall that persistence lengths
are scaled by the radius of the ring and, therefore, $\left\langle \delta
Wr^{2}\right\rangle \sim r^{2},$ in agreement with the scaling estimates in
reference \cite{Maggs00}. Indeed, writhe is a quadratic form of $\delta
\varphi $ and $\delta \theta $, each of which has typical fluctuations of $%
\sqrt{r/a}$, where $a$ is a characteristic persistent length and, therefore,
the characteristic amplitude of writhe fluctuations is $\delta Wr\simeq r/a$.

We proceed to calculate the probability distribution function of writhe, $%
P\left( \delta Wr\right) =\left\langle \delta \left[ \delta
Wr+\sum\nolimits_{n\neq 0,\pm 1}in\tilde{\varphi}(n)\tilde{\theta}(-n)\right]
\right\rangle .$ Using the exponential representation of the $\delta $%
-function, and performing the resulting integrations, a straightforward
calculation gives 
\begin{equation}
P\left( \delta Wr\right) =\sum_{n=2}^{\infty }\pi (n)%
%TCIMACRO{\dfrac{1}{2Wr(n)}}%
%BeginExpansion
{\displaystyle{1 \over 2Wr(n)}}%
%EndExpansion
\exp \left[ -%
%TCIMACRO{\dfrac{|\delta Wr|}{Wr(n)}}%
%BeginExpansion
{\displaystyle{|\delta Wr| \over Wr(n)}}%
%EndExpansion
\right] ,  \label{P4a}
\end{equation}
where 
\begin{equation}
\pi (n)=\left( -1\right) ^{n}%
%TCIMACRO{
%\dfrac{\pi n^{2}\left( n^{2}+\alpha \right) \left( 1+\alpha \right) \sqrt{\alpha }}{2\left( n^{2}-1\right) \sinh \left( \pi \sqrt{\alpha }\right) }}%
%BeginExpansion
{\displaystyle{\pi n^{2}\left( n^{2}+\alpha \right) \left( 1+\alpha \right) \sqrt{\alpha } \over 2\left( n^{2}-1\right) \sinh \left( \pi \sqrt{\alpha }\right) }}%
%EndExpansion
\quad \text{and \quad }\alpha \equiv 
%TCIMACRO{\dfrac{\left( n^{2}-2\right) a_{b}-a_{t}}{a_{b}+n^{2}a_{t}}}%
%BeginExpansion
{\displaystyle{\left( n^{2}-2\right) a_{b}-a_{t} \over a_{b}+n^{2}a_{t}}}%
%EndExpansion
.  \label{calc1}
\end{equation}
The above expression for $\pi (n)$ can be used to calculate all even moments
of writhe fluctuations (odd moments vanish because of the radial symmetry of
the undeformed ring), $\left\langle \delta Wr^{k}\right\rangle
=k!\sum_{n=2}^{\infty }\pi (n)Wr^{k}(n)$. The writhe distribution function
can be written in the form $P\left( \pi a_{b}\delta Wr,a_{b}/a_{t}\right) \ $%
and in Fig. 3 we plot this function vs. $\pi a_{b}\delta Wr$, for $%
a_{b}/a_{t}=0.1,$ $5$ and $20$. As expected, the probability of
large--amplitude writhe fluctuations (large $\left| \delta Wr\right| $)
decreases with increasing twist rigidity (decreasing $a_{b}/a_{t}$). The
width of the distribution decreases sharply when $a_{b}/a_{t}$ is of order
unity. With further increase of twist rigidity, the narrowing becomes less
pronounced and finally saturates for $a_{b}/a_{t}<0.1$. The distribution is
non--Gaussian, with an exponential tail at $\pi a_{b}\delta Wr\gg 1$. The
shape of the curve is qualitatively similar to recent computer simulations
\cite{Beard}, but detailed comparison could not be made since the
simulations were carried out in the $\delta Wr\gg 1$ regime (our study is
limited to the $\delta Wr\ll 1$ range).

In this letter we presented the statistical mechanics of fluctuating rings.
We derived analytical expressions for the two--point correlation functions
of the Euler angles, for the real space correlation functions and for the
probability distribution function of the writhe, as a function of the
persistence lengths associated with the bending and twist deformations of
the ring. We found that the amplitudes of fluctuations diverge in the limit
of vanishing twist rigidity and, therefore, theories in which only bending
rigidity is taken into account, can not be used to model fluctuating rings.
We found that a crossover length scale $\xi _{t}=r\sqrt{a_{t}/a_{b}}$
exists, below which writhe and twist fluctuations are decoupled, and above
which twist affects the three--dimensional configurations of the centerline
of the ring. It appears plausible that this crossover does not depend on the
topology of the ring and is characteristic of filaments with spontaneous
curvature in their stress--free state.

Finally, we would like to comment on the limitations of the present work.
Our theory is limited to the weak fluctuation regime, in the sense that the
deviations of the Euler angles from their equilibrium values must be
sufficiently small. In order to ensure this, the radius of the ring has to
be smaller than all the persistence lengths. For simplicity, we only
considered the case of a ring with a circular cross section. This assumption
is not essential and the study of a ring with asymmetric cross section and
three independent elastic constants, will be reported in an expanded version
of this letter, together with the study of linear response to externally
applied moments and forces, based on the fluctuation--dissipation theorem
that relates the correlation functions to the appropriate response functions
\cite{PRE}.

YR acknowledges support by a grant from the Israel Science Foundation.

\newpage

{\LARGE {\bf Figure captions}}

{\bf Fig. 1:} Plots of two--point correlation functions of Euler angles vs.
the dimensionless contour distance between the points, $s$, in the interval $%
0\leq s\leq 2\pi $: $\left\langle \delta \theta (s)\delta \theta
(0)\right\rangle $ (box), $\left\langle \delta \varphi (s)\delta \varphi
(0)\right\rangle $ (diamond), $\left\langle \delta \psi (s)\delta \psi
(0)\right\rangle $ (cross) and $\left\langle \delta \theta (s)\delta \psi
(0)\right\rangle $ (solid line). The bare persistence lengths are $%
a_{b}=a_{t}=10$.

{\bf Fig. 2:} Plot of real space two--point correlation function $%
\left\langle \left[ {\bf x(}s)-{\bf x(}0)\right] ^{2}\right\rangle $ vs. the
dimensionless contour distance between the points, $s$, in the interval $%
0\leq s\leq 2\pi .$ The parameters are $a_{b}=10$, $a_{t}=$ $10$ (solid
line), $a_{b}=10$, $a_{t}=1$ (box) and $a_{b}=1$, $a_{t}=$ $10$ (cross).

{\bf Fig. 3:} Plot of probability distribution function of writhe vs. $\pi
a_{b}\delta Wr$ for $a_{b}/a_{t}$ $=$ $0.1$ (solid line), $5$ (cross) and $%
20 $ (box).

\end{document}